\documentstyle[11pt,newpasp,twoside,epsf]{article}
\def\edcomment#1{\iffalse\marginpar{\raggedright\sl#1\/}\else\relax\fi}
\reversemarginpar

\begin{document}
\title{The Red-Sequence Cluster Survey}
\author{Michael D. Gladders}
\author{H.K.C. Yee}
\affil{Department of Astronomy, University of Toronto,
60 St. George Street, Toronto, ON, Canada, M5S 3H8}

\begin{abstract}
The Red-Sequence Cluster Survey (RCS) is a new galaxy cluster survey
designed to provide a large sample of optically selected 0.1$<$z$<$1.4
clusters. The planned survey data are 100 square degrees of two color
($R$ and $z'$) imaging, with a 5$\sigma$ depth $\sim$2 mag past
M$^{*}$ at $z=1$. The combined depth and area of the RCS make it the
widest field, moderately deep survey ever undertaken using 4m class
telescopes.  This paper gives a brief outline of the RCS survey, with
particular emphasis on the data reduction strategy.  The remainder of
the paper focuses on preliminary results from the first set of
completely reduced data ($\sim$10 deg$^2$, of the $\sim$60 deg$^2$ in
hand). We provide a new example of a rich $z>1$ cluster, illustrative
of the dozens discovered in the data so far. Some of the possible
science to come from the RCS is illustrated by a qualitative
indication of $\Omega_{M}$ from the first 1/10th of the survey
data. A high-precision measurement of the 2-point correlation function
of luminous early-type galaxies at $0.4<z<1.2$ is also shown.
\end{abstract}

\section{Introduction}
\paragraph{}
Galaxy clusters are the largest gravitationally bound structures in
the universe, and as such are markedly important in cosmology and
studies of galaxy evolution. Much of the work on clusters done to date
at low to intermediate redshifts (typically limited to $\sim0.8$ or
so) would benefit greatly from extension to higher redshifts. However,
the cluster samples required for such studies do not exist. The
Red-Sequence Cluster Survey (RCS) is designed to rectify this problem.
The survey is designed to find galaxy clusters in large numbers out to
$z\sim1.4$, spread over a wide range of RA and DEC in order to
facilitate follow-up.

\paragraph{}
The first observations for the RCS were taken in May 1999, at the
CFHT, and at the time of writing we have acquired $\sim60\%$ of the
100 square degrees of planned survey data. Data acquisition will be
completed by the end of 2001. The survey data consist of 22 survey
patches (ten $2.1\times2.3$ deg$^2$ patches from the CFHT, and twelve
$1.8\times2.4$ deg$^2$ patches from the CTIO 4m telescope) imaged
in the $R_C$ and $z'$ filters to $5\sigma$ point source depths of 25.2
and 23.6 mags respectively. This depth has been chosen to facilitate
finding $z\sim1$ galaxy clusters, which are located in the survey data
using an algorithm based on detecting the early-type galaxies in the
clusters cores (Gladders \& Yee 2000a). The RCS has been described in
more detail elsewhere (Gladders \& Yee 2000b), and we will not provide
a repetition here. The interested reader is referred to the survey
homepage, {\texttt {http://www.astro.utoronto.ca/$\sim$gladders/RCS/}}, which
contains a description of the survey strategy, goals, and other
relevant data and publications.
\paragraph{}
The purpose of this paper, rather, is to provide some of the details
of the survey design strategy, as these are germane to the topic of
this meeting. In addition, we provide some examples of new results
coming from the RCS, including a preliminary indication of $\Omega_M$,
an example of $z\sim1$ galaxy clusters, and some of the secondary
science related to the identification of individual early-type
galaxies using photometric techniques over an unprecedented volume to
$z\sim1$.
\section{RCS Survey Strategy}
\paragraph{}
The primary science driver of the RCS is to locate a large sample of
high redshift galaxy clusters, suitable both for direct investigations
with the survey data and detailed follow-up work. The need for
detailed follow-up at these redshifts is mandated by the paucity of
known high-redshift galaxy clusters. At a redshift of $\sim0.8$ and
beyond, we know almost nothing about such structures. We thus chose to
complete the survey in a large number of (relatively) small patches
spread over a wide range in RA and DEC. This basic design means that
the RCS sample will be observable from $all$ major observatories at
$all$ times of year.
\paragraph{}
In considering the survey design, it was also recognized that the
primary difficulty in producing such a survey is not the telescope
time (the entire survey requires only 25 clear nights, split between
two telescopes) but rather the ability of the survey team to handle
the data flow. Consider, for example, that the CFHT half of the survey
consists of $\sim150$ CFH12K images (in two filters), or a total of
150$\times$12$\times$2=3600 2k$\times$4k CCD final science
images! This large data flow along with the relatively short exposure
times led to the decision to acquire all the data without
dithering. This greatly simplifies the construction of final science
images, at the expense of some minor loss in area due to inter-chip
gaps and cosmetic defects. Cosmic rays, the removal of which is often
cited as a reason for dithering, are accounted for in the photometry
pipeline. The use of undithered images means that the entire survey
can be treated as an assemblage of individual 2k$\times$4k images, for
which the astrometric and photometric calibration can be performed
solely in the final catalog assembly stage.
\paragraph{}
The survey data reduction pipeline works as follows. Individual chips
from the mosaic imagers are pre-processed (de-biassed, flat-fielded,
de-fringed etc.) using standard techniques. Object finding, photometry
and star-galaxy seperation are then done using an updated version of
PPP (Yee 1991). For each chip, a preliminary catalog is produced from
the object list with no visual checking. A final catalog results once
the object finding has been visually checked, an unavoidably
time-consuming process which removes the few defects which are
impossible to filter out automatically. The catalogs are then stitched
together into a patch catalog using astrometric information from the
USNO-A2.0 Catalog and photometric standards from Landolt (1992) and
Thuan \& Gunn (1976). The RCS will eventually be put onto the Sloan
Digital Sky Survey photometric system (Fukugita et al. 1996); this
calibration awaits analysis of data now in hand, and publication of the
SDSS photometric calibration. Internal cross-checks of the photometry
using 30$''$ overlap regions between each pointing in a patch shows
that the preliminary calibration has an internal accuracy of better
than 0.05 mags.
\section{Preliminary Science Results}
The results shown below are drawn primarily from the first $\sim10$
deg$^2$ of RCS data. The various analyses should be considered
preliminary, but are indicative of the overall significance of the survey.
\subsection{Cosmology} 
The redshift evolution of the mass spectrum of galaxy clusters,
$N(M,z)$, provides a strong measure of the cosmological parameters
$\Omega_M$ and $\sigma_8$. Analysis of the RCS in this context will be
made, at a later stage, using richness as a measure of mass (Yee \&
Lopez-Cruz 1999) and photometric redshifts. As a qualitative
illustration of the cosmology indicated by the RCS, we have used the
RCS selection functions (Gladders \& Yee 2000a,c) to integrate out
the mass dependence in $N(M,z)$ and compared the measure of $N(z)$ to
predictions from two typical cosmologies. Predictions of $N(M,z)$ are
made from the standard Press-Schechter formalism, and multiplied by
the RCS selection functions (expressed in mass and redshift, assuming
a cluster with a typical luminosity function shape, galaxy mixture and
cluster shape, concentration and size). These selection functions tail
off to zero probability at a lower mass limit which becomes
progressively more massive at higher redshift and hence limit the
contribution to $N(M,z)$ from lower-mass clusters and groups.  The
result of these computations, as well as the actual counts from a
portion of the RCS, are shown in Figure 1. Clearly, and not
suprisingly, the low $\Omega_M$ and high $\sigma_8$ model is
preferred.

\begin{figure}[!h] 
\plotfiddle{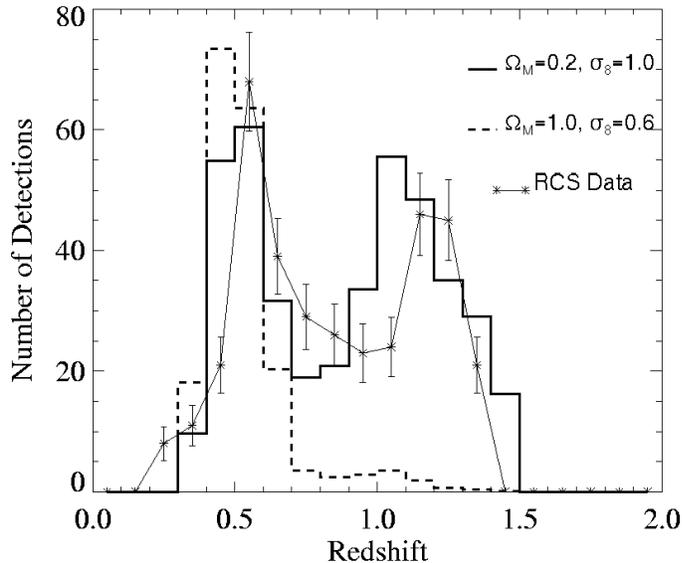}{7cm}{0}{50}{50}{-175}{-100}
\caption{An illustration of the cosmological power of the RCS. The
predicted counts for clusters and groups are shown for two different
cosmologies (thick lines) as well as the measured counts of
significant detections in a subset of the RCS (thin line + asterisk).
The low $\Omega_M$ model is vastly preferred. The apparent excess of low
redshift objects (resulting in an apparent `double-peaked' appearance)
is the result of redshift aliasing from lower redshifts, and the great
sensitivity of the RCS to even $300$ km~s$^{-1}$ groups at moderate
redshifts (Gladders \& Yee 2000a).  }
\end{figure}

\subsection{High Redshift Clusters}
Example high redshift clusters from the RCS have been shown elsewhere
(e.g., Gladders \& Yee 2000b), and we will not repeat those illustrations
here. Instead, Figure 2 shows one of a new set of clusters found in an
analysis of $\sim20$ deg$^2$ of preliminary RCS catalogs. The
photometric redshift is in the range $z>1.3$, and this is the single
richest high redshift cluster so far discovered in the RCS. 

\begin{figure}[!h] 
\plotfiddle{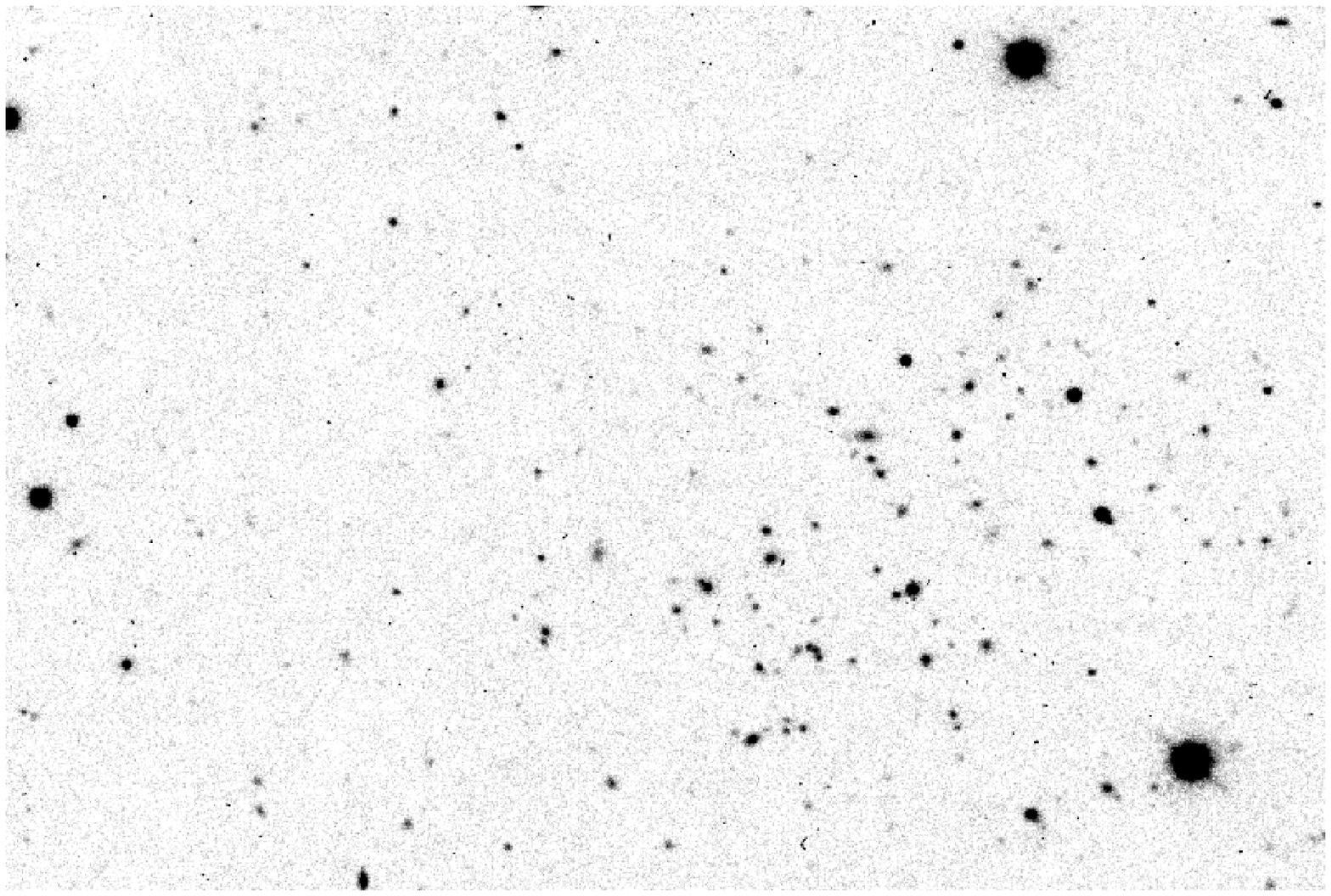}{7cm}{0}{60}{60}{-190}{-130}
\caption{A $z>1.3$ rich cluster from the RCS. The image is about
$3\times2$ arcmin in size, in the $z'$ filter. The cluster is the
concentration of galaxies to the right. A $R_Cz'$ color image
can be found at the RCS homepage.}
\end{figure}

\subsection{The 2-Point Correlation Function of Early-Type Galaxies}
\paragraph{}
Over the redshift range in which the RCS filter pair straddles the
4000\AA~~break ($0.4<z<1.2$), the bright red edge of the
color-magnitude diagram for all galaxies must be dominated by
early-types. Moreover, the color is a direct measure of the
redshift. This is because, at a given color, all lower redshift galaxies
appear bluer, all higher redshift early-type galaxies appear redder, and
all later-type higher redshift galaxies (which may have the same apparent
color) appear fainter. Thus, at the bright red edge, at a given color,
the sample is dominated almost exclusively by early-type galaxies at a
certain redshift. It is thus possible to isolate a sample of
luminous early-type galaxies simply by making a magnitude-color cut which
selects the bright red edge of the color-magnitude diagram.
\paragraph{}
We have tested this idea by performing such cuts in the CNOC2 (Yee et
al. 2000) database, which provides both redshifts and spectral-energy
distributions (a surrogate for morphological classification). Using
$V$ and $I_C$ data from CNOC2, and a passively evolving $z=3$
starburst model to predict elliptical galaxy colors and apparent
magnitudes, we find that a cut to a depth of $M^*+1$ provides an
early-type galaxy sample which is minimally contaminated
($\sim15\%$), and for which individual redshifts are
estimated to an accuracy of $\Delta z=0.05$.
\paragraph{}
Using this technique, we have isolated a sample of $\sim10,000$
early-type galaxies from the RCS in the redshift range $0.4<z<1.2$,
from an area of $\sim6$ deg$^2$.  The sample was split into redshift
bins, and the angular two-point correlation function of each bin was
computed. Several example correlation functions are shown in Figure
3. 
\begin{figure}[!h] 
\plottwo{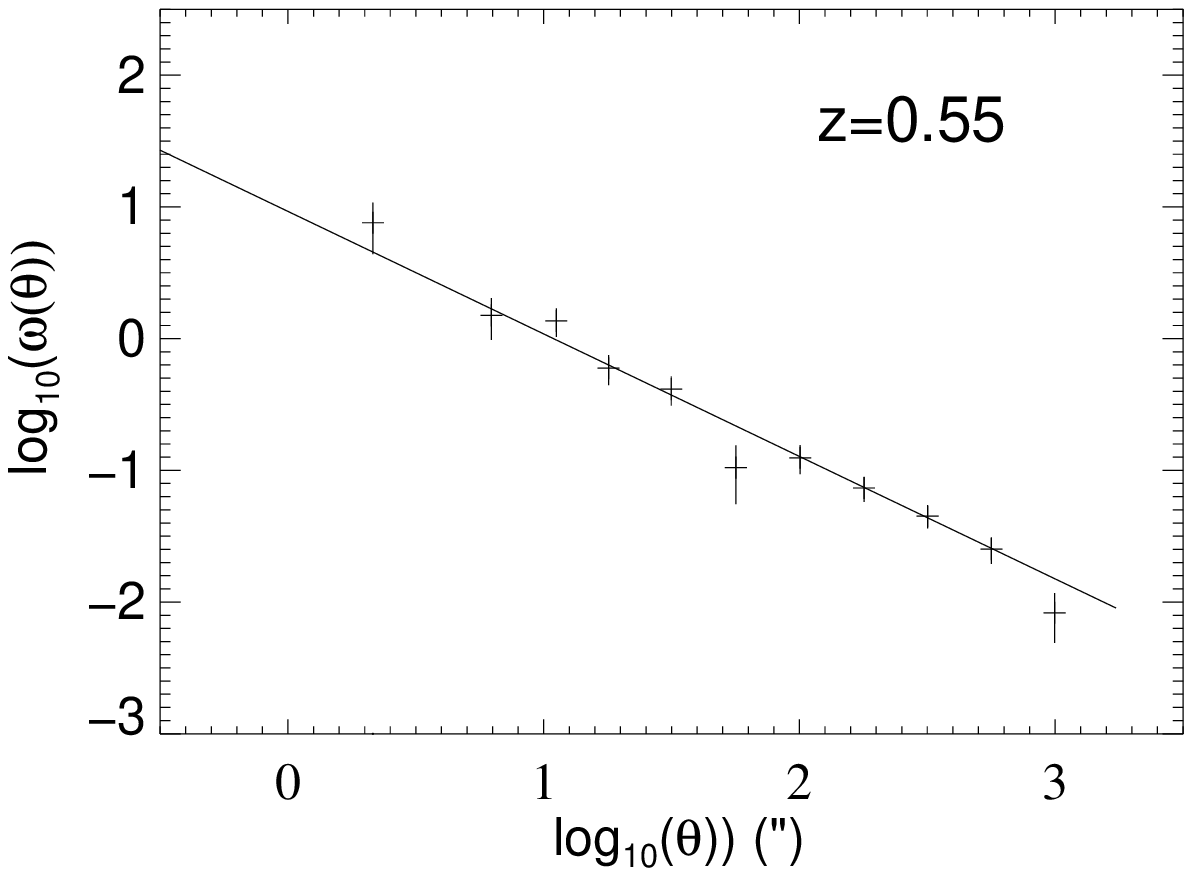}{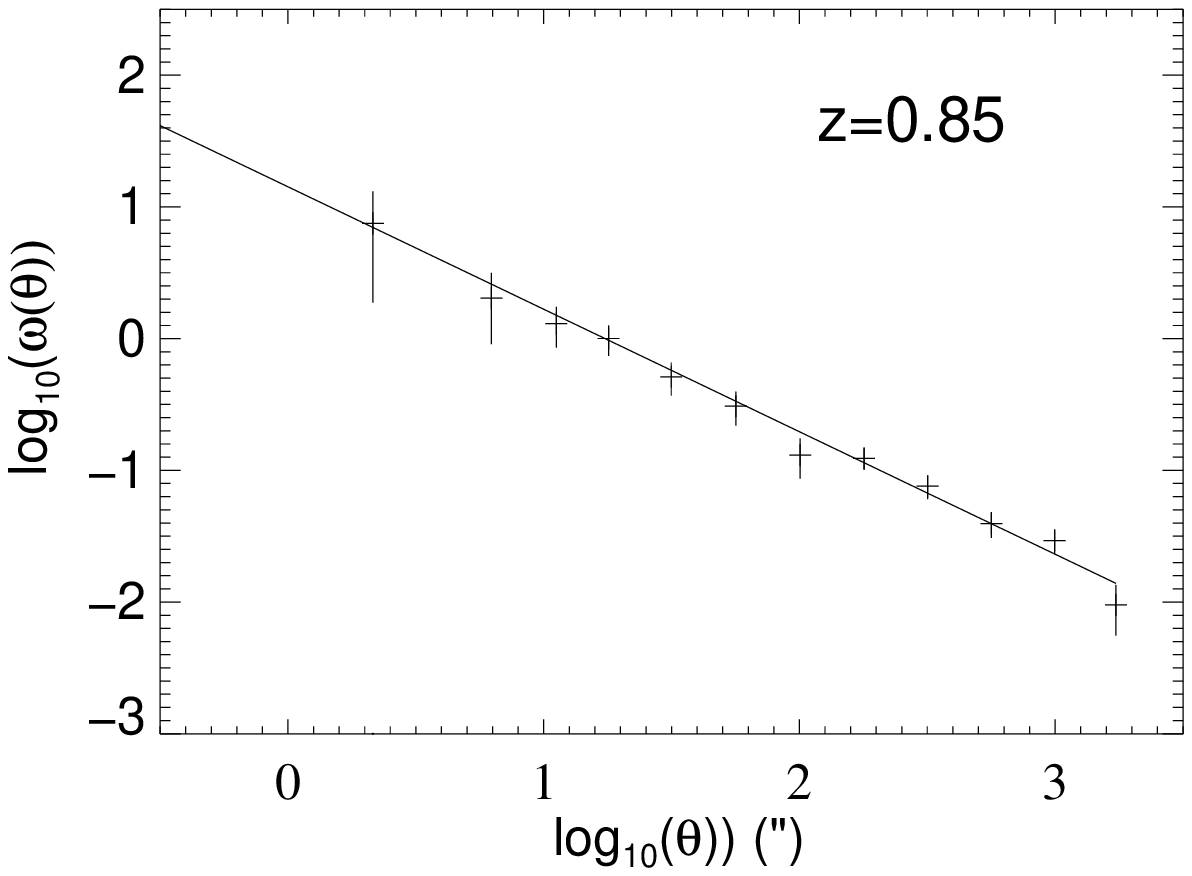}
\caption{Measured angular two-point correlation functions for luminous
early-type galaxies in two redshift bins centered at $z=0.55$ (left)
and $z=0.85$ (right). The bins have a width of 0.1 in redshift.}
\end{figure}
The measured angular correlation was then inverted via the Limber
transform (e.g., see Giavalisco et al. 1998) to deduce the physical
correlation length. Figure 4 shows these derived correlation lengths,
in addition to those recently reported from the CNOC2 redshift survey
by Shepherd et al. (2000). These measurements, as well as tests of the
robustness of the inversion, are presented in detail in a paper now in
preparation (Gladders \& Yee 2000d). Note in Figure 4 that, according
to this analysis, the correlation length for luminous early-type
galaxies is unchanged out to at least redshift one.

\begin{figure}[!h] 
\plotfiddle{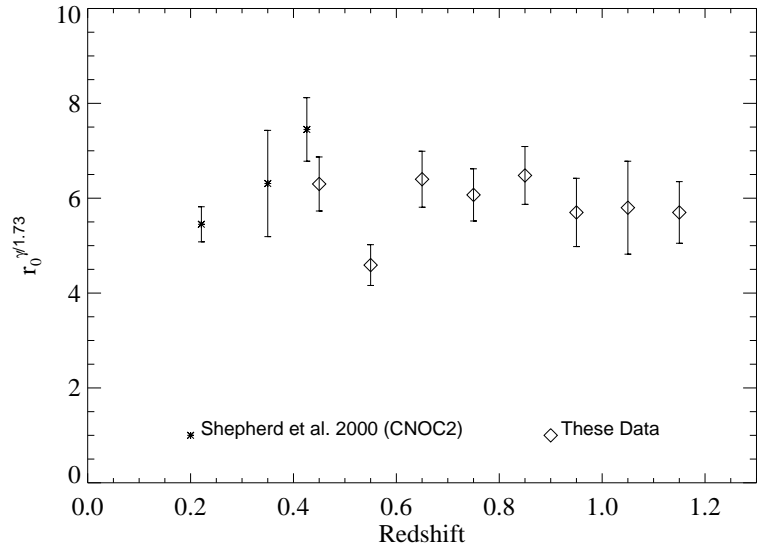}{7cm}{0}{40}{40}{-170}{0}
\caption{The measured co-moving correlation length of luminous
early-type galaxies from the RCS (diamonds). Also shown are the CNOC2
derived measurements of Shepherd et al. (2000) (asterisks). Both are
expressed relative to correlations lengths with a slope of
$\gamma=1.73$ (see Shepherd et al. for details).}
\end{figure}


\begin{references}
\reference Fukugita, M., Ichikawa, T., Gunn, J.E., Doi, M., Shimasaku, 
K., \& Schneider, D.P. 1996, AJ, 111, 1748
\reference Giavalisco, M., Steidel, C.,
Adelberger, K.L., Dickinson, M.E., Pettinin, M., \& Kellog, M. 1998,
ApJ, 503, 543
\reference Gladders, M.D., \& Yee, H.K.C. 2000a, AJ, 120, 2148
\reference Gladders, M.D., \& Yee, H.K.C. 2000b, to be published in 
ASP proceedings of {\it Cosmic Evolution and Galaxy Formation: Structure, 
Interactions and Feedback}, see astro-ph\/0002340
\reference Gladders, M.D., \& Yee, H.K.C. 2000c, to be submitted to AJ
\reference Gladders, M.D., \& Yee, H.K.C. 2000d, to be submitted to ApJ
\reference Landolt, A.U. 1992, AJ, 104, 372 
\reference Shepherd, C.W. et al. 2000, submitted to ApJ
\reference Thuan, T.X., \& Gunn, J.E. 1976, PASP, 88, 543
\reference Yee, H.K.C, \& L\'{o}pez-Cruz, O. 1999, AJ, 117, 1985
\reference Yee, H.K.C., Morris, S.L., Lin, H., Carlberg, R.G., Hall, P.B.,
Sawicki, M., Patton, D.R., Wirth, G.D., Ellingson, E., \& Shepherd,
C.W. 2000, ApJS, 129, 475


\end{references}
\end{document}